\documentclass{llncs}

\usepackage{epsfig}
\usepackage{epstopdf}
\usepackage{pdfpages}
\usepackage{graphicx}
\usepackage{url}
\usepackage{changebar}
\usepackage{xcolor}
\usepackage{soul}
\usepackage[caption = false]{subfig}
\usepackage{url}
\usepackage{xspace}

\newcommand{\todo}[1]{}
\newcommand{\MPI}[1]{\mbox{MPI-#1}\xspace}

\newif\ifdiff
\difftrue 

\ifdiff
\newcommand{\removed}[1]{\cbstart\removedfragile{#1}\cbend{}}
\newcommand{\removedfragile}[1]{{\color{red}\st{#1}}{}}

\else
\newcommand{\removed}[1]{} 

\fi

\begin{document}
\title{Leveraging MPI-3 Shared-Memory Extensions for Efficient PGAS
  Runtime Systems}

\author{
  Huan Zhou
  \and Kamran Idrees
  \and Jos\'e Gracia
}
\institute{
  High Performance Computing Center Stuttgart (HLRS), University of Stuttgart, Germany
}

\maketitle

\begin{abstract}
  The relaxed semantics and rich functionality of one-sided
  communication
  primitives of \MPI3 makes MPI an attractive candidate for
  the implementation of PGAS
  models. However, the performance of such implementation suffers
  from the fact, that current MPI RMA implementations 
  typically have a large overhead when source and target of a
  communication request share a common, local physical memory.  In
  this paper, we present an optimized PGAS-like runtime system
  which uses the new \MPI3 shared-memory extensions to
  serve intra-node communication requests and \MPI3 one-sided
  communication primitives to serve inter-node communication requests.
The performance of our runtime system is evaluated on a Cray XC40
system through low-level communication benchmarks, a random-access
benchmark and a stencil kernel.  The results of the experiments
demonstrate that the performance of our hybrid runtime system matches
the performance of low-level RMA libraries for intra-node transfers,
and that of \MPI3 for inter-node transfers.
\end{abstract}
\begin{keywords}
 MPI, one-sided communication, remote-memory access, RMA, partitioned
 global address space, PGAS
\end{keywords}

\section{Introduction}
\todo{Why only blocking: direct memcopy is a
blocking operation, only. Therefore need progress engine or similar
for non-blocking hybrid.}
The Message Passing Interface (MPI, \cite{mpi3}) is the de-facto
communication standard for distributed-memory parallel programming.
One particular advantage for parallel programmers is the portability
of MPI performance across systems with different underlying network
hardware: While HPC hardware vendors and the MPI community spend
considerable effort to optimize MPI implementations for the latest HPC
network infrastructure, other alternative communication libraries
typically do not have optimized support for a wide range of network
hardware.
With the advent of the remote-memory access (RMA, also referred to as
one-sided communication) functionalities in \MPI2\cite{mpi22} and the
significant improvement of the RMA in \MPI3\cite{mpi3}, MPI has become an
adequate communication backend for the implementation of partitioned
global address space (PGAS) programming models\cite{pgas}.

DASH\cite{dash} is a C++ template library which implements a PGAS-like
programming model. Unlike other PGAS models, DASH acknowledges the
multi-level hierarchical or compositional nature of today's     
supercomputing systems, e.g cores, processors, nodes, racks, islands,
full system, and thus does not classify data into remote and local
only, but allows for various degrees of remoteness. The template
library sits on top of a runtime system (DART), which is responsible
for providing services to the DASH library, including the definition
of semantics and the abstraction of the underlying hardware. In
particular, DART provides functions for the management of teams
(a concept similar to  MPI communicators),
one-sided communications, collective operations,
and global memory management. 

In an earlier paper\cite{dart-mpi}, we have described DART-MPI, a portable
implementation of the DASH runtime, that uses \MPI3 as low-level
communication substrate. There, we showed, that the overhead of
DART-MPI RMA operations on top of the corresponding MPI-3 operations
is negligible in general. Most other PGAS
implementations however, do not use MPI as communication substrate;
UPC\cite{upc} for instance is frequently based on
GASNet\cite{techgasnet}, while GA\cite{ga} uses ARMCI\cite{armci} as
underlying communication substrate.

Originally, all the RMA operations in DART-MPI are substantially mapped
directly to the corresponding \MPI3 RMA operations. In particular,
DART-MPI invokes MPI RMA operations when source and target of a
transfer reside on the same node and share local,
physical memory. Alternatively, one could do direct load/store
operations without additional copies in the runtime layer. 
In this paper, the contributions we make on \mbox{DART-MPI} are threefold:
\begin{itemize}
  \item We utilize the \MPI3 shared-memory extensions to 
    enable direct memory access (memory sharing)
     for \mbox{DART-MPI} blocking operations for intra-node transfers. 
    However, we turn to 
    the MPI RMA operations when the non-blocking or inter-node data movements happen.
\item We redefine the existing translation table to facilitate the reference to the DART-MPI
collective global pointer
when beginning with the shared memory window in mind.
\item Using the low-level and application-level benchmarks, we show the improved performance
achieved by embedding the shared-memory-related functionality into \mbox{DART-MPI}.
\end{itemize}

\todo{Rest of paragraph into discussion?!\\
This work has demonstrated
that the shared-memory functionality proposed in the \MPI3
benefits the \mbox{DART-MPI} intra-node blocking RMA operations. However, \mbox{DART-MPI} turns
to the MPI RMA operations for the non-blocking or inter-node scenario,
leading to low performance.}

The rest of the paper is organized as follows: In Sect. \ref{background}, we present the background for our work. In Sect. \ref{design},
we describe the improved implementation of \mbox{DART-MPI} and evaluate the performance of \mbox{DART-MPI} in Sect. \ref{sec:eval}. We summarize in Sect. \ref{conclusion}.

\section{Background}
\label{background}
From the perspective of PGAS models, the recent \MPI3 standard
\cite{mpi3} significantly improves the one-sided communication
system. The relaxation of the RMA semantics, the
concretization of the memory consistency model, the introduction of
new window types, fine-grained mechanisms for
synchronization 
and data movement, and atomic operations
, make MPI-3 RMA attractive as backend for PGAS implementations. Additionally, the results
in Dinan et al.~\cite{Dinan} indicate that the new \MPI3 RMA system has
performance advantages over the \MPI2 interface.  In this section, we
briefly explain two new \MPI3 window types:
\emph{dynamically-allocated window} and \emph{shared-memory window},
which will play a central role in understanding how to enable memory sharing
within a node in \mbox{DART-MPI}.  A more detailed description
of the other new functionalities can be found in Hoefler et al.~\cite{Hoefler}.

\subsection{MPI Dynamically-allocated Memory Window}
A dynamically-allocated window is a new concept in \MPI3 that allows to
arbitrarily grow and resize a given window by repeatedly attaching/detaching multiple, 
non-overlapping, user allocated memory regions to/from the associated 
window object.

The function \textit{MPI$\_$Win$\_$create$\_$dynamic} is called
to generate a window object \textit{d-win} without associating any
initial memory block with it. User allocated memory is attached to
\textit{d-win}, and thus made available for RMA operations, by
invoking the function \textit{MPI\_Win\_attach}, and detached with
\textit{MPI\_Win\_detach}.  Once memory regions are detached from
\textit{d-win}, they will not be the target of any MPI RMA operation
on \textit{d-win} unless they are re-attached. Notably, any local
memory region may be attached and detached repeatedly, and multiple,
but \mbox{non-overlapping} memory regions are allowed to be attached
to the same window. 

{\em MPI$\_$Get$\_$address} returns the address of 
the given memory and should be called to validate the RMA operations on \mbox{\textit{d-win}}.
This is due to the fact that the address of the target memory location is passed directly
as window displacement parameter to the MPI RMA operations.
Therefore, the target process is required to send the address of a certain memory location, 
that locals to it, to the origin process who inquires for it.

Noticeably,  Potluri et al.\cite{dynamic} have published benchmark
results which demonstrate that dynamically-allocated
windows perform as good as the traditional static MPI-created windows in terms of 
put latency.

\subsection{MPI Shared-memory Window}
The unified memory model, which is fully 
supported in \MPI3 in order to utilize the \mbox{cache-coherence} characteristics embodied in the modern hardware
architectures, is a requirement for exposing the MPI \mbox{shared-memory} window.

To collectively allocate 
the shared memory region across all processes in a given communicator,
\MPI3 defines a portable, \mbox{shared-memory} window allocation interface -- {\em MPI$\_$Win$\_$allocate$\_$shared}
to generate a \mbox{shared-memory} allocated window object \mbox{\textit{shmem-win}}.
In addition, the communicator that the \mbox{\textit{shmem-win}} associates
with should be a \emph{shared-memory capability} 
communicator, which means it is allowed to build a memory sharing region on top of this communicator.
Therefore, the additional function {\em MPI$\_$Comm$\_$split$\_$type}, 
as an extension of the function {\em MPI$\_$Comm$\_$split}, identifies \mbox{sub-communicators} on which
the shared memory region can be created with the type of {\em MPI$\_$COMM$\_$TYPE$\_$SHARED}. 
The function {\em MPI$\_$Win$\_$shared$\_$query} is provided to query the base pointer to 
the memory on the target process. Coupled with the \mbox{\textit{shmem-win}}, 
the locally-allocated memory can even be accessed by the MPI processes in the group of
\mbox{\textit{shmem-win}} with immediate
load/store operations. Such access pattern can make data movements bypass the 
MPI layer and directly go through memory sharing, which brings in significant performance improvement.

\section {The \mbox{DART-MPI} Implementation Design}
\label{design}

In this section, 
we explain the approach of enabling the memory sharing option for the blocking RMA operations in \mbox{DART-MPI}
and address the modifications and improvements that are made with respect to the existing \mbox{DART-MPI}.

There are two types of DART global memory, collective and non-collective\cite{dart-mpi}.
The collective global memory, pointed to by a collective global pointer, 
is created and distributed across the given team.
The non-collective global memory, pointed to by a non-collective global pointer, 
is only allocated in the global address space of the calling unit.
We assume that all the following collective global memory blocks are
allocated across team $T$ consisting of $P$ units.

\subsection{Communication Hierarchy of the DART-MPI Blocking RMA Operations}
To make the DART-MPI \mbox{intra-node} communication more efficient,
we alter the existing implementation
to let the \mbox{DART-MPI} blocking operations deal with the data locality explicitly.
\todo{Following sentence in discussion?} Note, that the DART-MPI non-blocking RMA interfaces do not yet support the memory sharing as 
described earlier in this paper. 

In the team creation code, 
the team $T$ is split into \mbox{sub-teams} on which it is possible to enable communication via sharing memory. 
We accomplish this by calling {\em MPI\_Comm\_split\_type} with key
{\em MPI\_COMM\_TYPE\_SHARED}. In addition, a \mbox{\textit{d-win}} 
is generated without any memory attached 
when team $T$ is created, indicating \mbox{one-to-one} relationship is built between 
\mbox{\textit{d-win}} and $T$. Such relationship is stored in an array named {\em dart$\_$win$\_$lists}.
Therefore, the position of the team $T$ in \textit{teamlist}\cite{dart-mpi} can also be a perfect index into the 
array {\em dart$\_$win$\_$lists}.
The \mbox{\textit{d-win}} can potentially be utilized to complete
all the data movements where the units are located in different \mbox{sub-teams}.

\begin{figure}[tb]
\begin{center}
\includegraphics[scale=0.40]{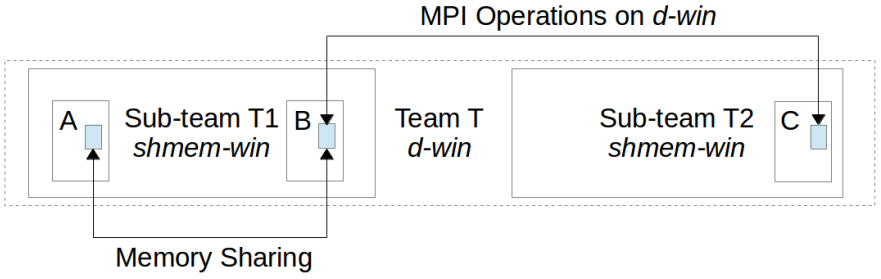}
\end{center} 
\caption{Nesting of shared-memory window inside RMA window for blocking put/get operations}
\label{structure}
\end{figure}

In the collective global memory allocation code, instead of allocating a block of memory 
from a memory pool that is reserved for $T$, we need to create a \mbox{\textit{shmem-win}} spanning the 
memory of the specified size on each sub-team mentioned above.
On top of that, each unit of the team $T$ should attach the locally-allocated memory
to the \mbox{\textit{d-win}} explicitly to make them available for the units in the varying \mbox{sub-teams}. As the 
Fig. \ref{structure} shows, there are two overlapping windows sharing the same memory region for different purposes.
On the one hand, the units covered by the same \mbox{\textit{shmem-win}} can communicate with each other via
memory sharing (e.g., memcpy). On the other hand, the units located in different \mbox{sub-teams} should turn to the \mbox{\textit{d-win}} for
completing the remote accesses with MPI RMA operations.
Note, that using shared-memory in the 
DART-MPI non-blocking RMA operations is anything but trivial as for instance the direct memcpy function itself 
is a blocking operation. Furthermore, the introduction of DMA copy engine could be a workaround to support 
asynchronous memory copying~\cite{async} for DART-MPI. 

In the \mbox{non-collective} global memory allocation code, we provide two overlapping global 
windows, 
which indicates that all the \mbox{DART-MPI} \mbox{non-collective} 
allocations fall into two \mbox{pre-defined} global windows. 
One of the two windows is generated first spanning a large amount of shared memory region 
on the default communicator -- {\em MPI\_COMM\_WORLD}~\cite{mpi3} for intra-node 
communications, the other is
then created with {\em MPI$\_$Win$\_$create} covering the above
shared memory region to enable the message transferring across different nodes. As a result,
these two windows share the same static shared memory region, and independently 
implement the data movements on them in an efficient manner.
\subsection{DART-MPI Collective Global Pointer Dereference}
In this section, we 
mainly explain the collective global pointer dereference of 
the updated \mbox{DART-MPI} since the \mbox{non-collective} global 
pointer basically continues to use the original dereference mechanism.

Besides the altered communication pattern, the meaning of the member \textit{segid} in the global pointer
is also \mbox{re-specified}
for management convenience and data access efficiency. Therefore, the \textit{segid} 
in the collective global pointer is no longer set to the related team ID but rather
an increasing positive integer number, which 
can be used to determine any collective global block uniquely. 

With the aid of the translation table\cite{dart-mpi}, 
collective global pointer can get analyzed adequately.
Thus it is critical for us to understand how the translation table 
reacts to the hierarchical communication pattern and 
the modification made in the global pointer, which also
has an impact on the original collective global pointer deference method.

To be consistent with the modified definition of \textit{segid} in global pointer,
the key in the translation table is altered and the \textit{segid} is utilized instead.
The translation table is arranged in an ascending order based on the key \textit{segid}. 
As a result,
we do not need to bind a separate translation table to each team, 
instead a single translation table is active during the lifetime of a \mbox{DART-MPI} program. 
Once a block of collective global memory is created, a unique \textit{segid} 
and the related \textit{shmem-win} are generated
and then added to the translation table
together, signifying the \mbox{one-one} relationship between collective global pointer and the related \textit{shmem-win}.
In addition, according to the Sect. \ref{background}, we learn that after 
attaching the shared memory region onto the \mbox{\textit{d-win}} locally, the routine  
{\em MPI$\_$Get$\_$address} should be invoked
so as to collect the beginning address of the local shared memory region
of each unit in team $T$. Thus, the translation 
table should also contain an array \mbox{\textit{disp-set}}
storing those separate addresses. 
As an example, when unit $i$ in the team $T$ is targeted, then the $i$th item in the related
\mbox{\textit{disp-set}} should be obtained and be utilized in the future to locate the target memory location in unit $i$.
The offset returned in the generated 
collective global pointer is initialized to $0$.

The location of target data is given by DART global pointer, which incorporates the information 
on the target unit, 
\textit{segid} and a specific offset. 
For the collective global pointer, in the case of intra-node communications, 
we firstly query the appropriate \mbox{\textit{shmem-win}} that covers the expected target 
location from the translation table according to the \textit{segid}, then decode the location with 
offset. In the case of inter-node communications,
we firstly query the \mbox{\textit{disp-set}}, indicates the beginning address of the window segment of each unit in team $T$,
from the translation table according to the \textit{segid}, and then get the correct \mbox{\textit{d-win}} from the array {\em dart$\_$win$\_$lists}
and translate the absolute unit id to the relative unit id $i$ in $T$,
and finally access the remote data through MPI RMA operations, where
the value of \textit{offset}+\textit{disp-set}[i]
is passed as parameter \textit{target\_disp}.

\section{Performance Evaluation}\label{sec:eval}
\begin{figure}[bp]
\begin{center}
\subfloat[Blocking put (intra-node)]{\includegraphics[width=0.48\textwidth,height=0.21\textheight]{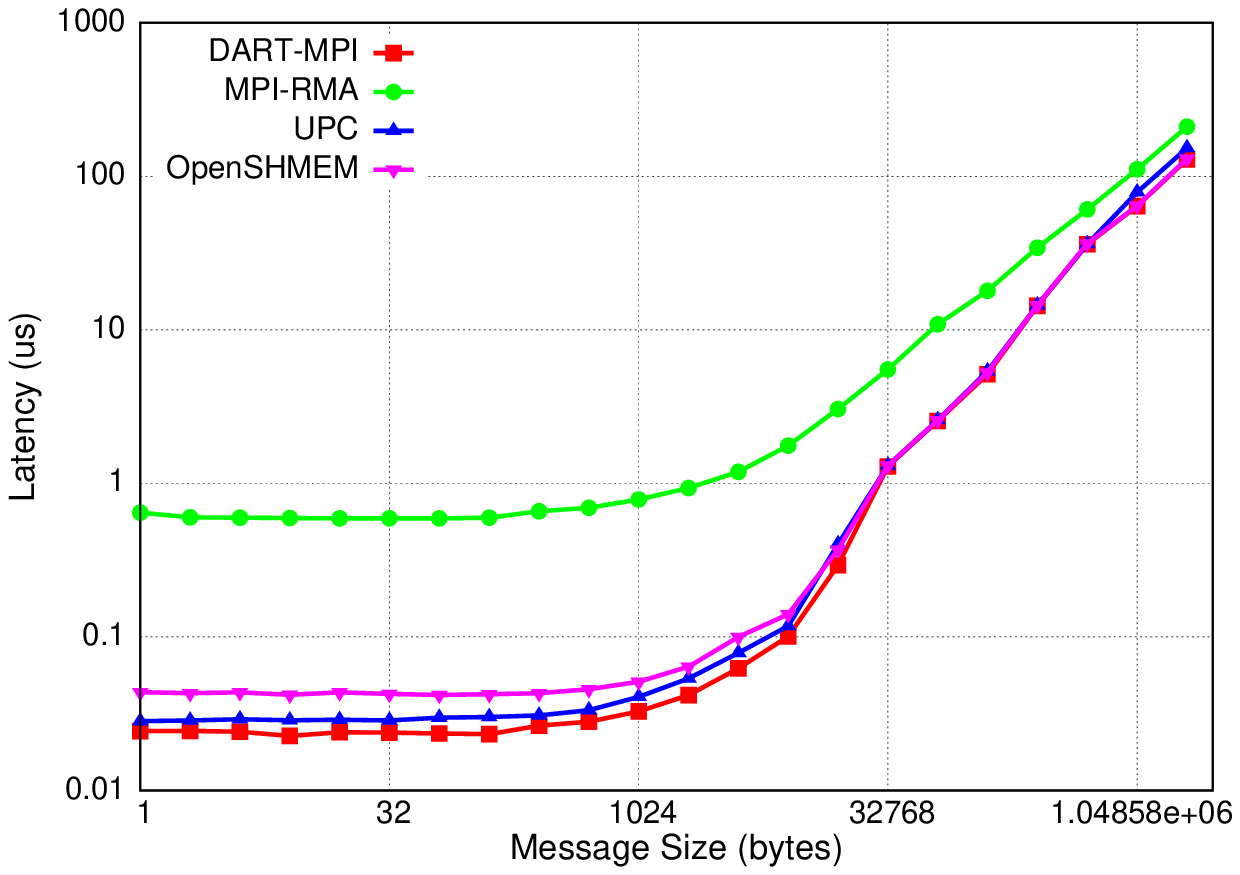}} 
\subfloat[Blocking get (intra-node)]{\includegraphics[width=0.48\textwidth,height=0.21\textheight]{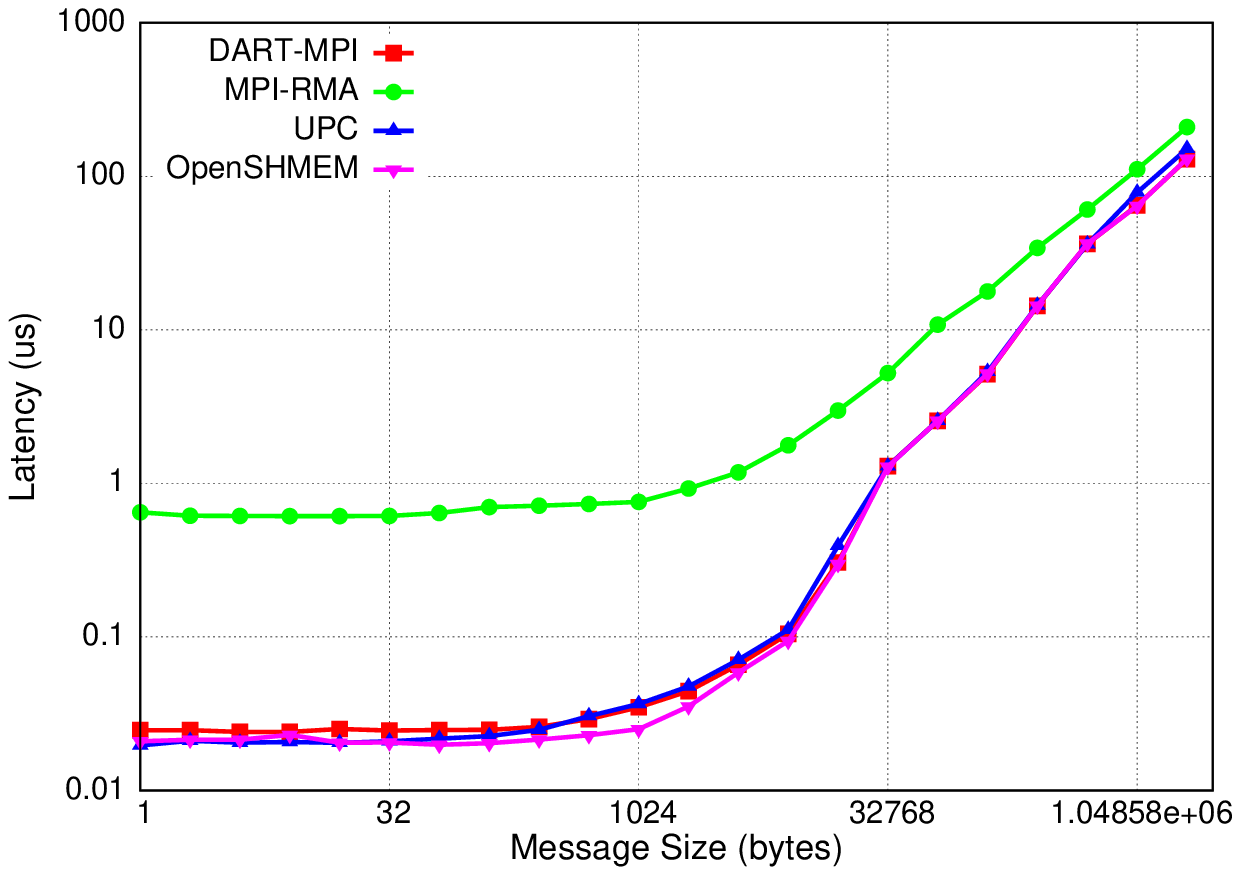}}
\end{center} 
\caption{Blocking put/get latency on 2 ranks/units}
\label{latency}
\end{figure}
\begin{figure}[tbp]
\begin{center}
\subfloat[Blocking put (8 bytes)]{\includegraphics[width=0.48\textwidth,height=0.21\textheight]{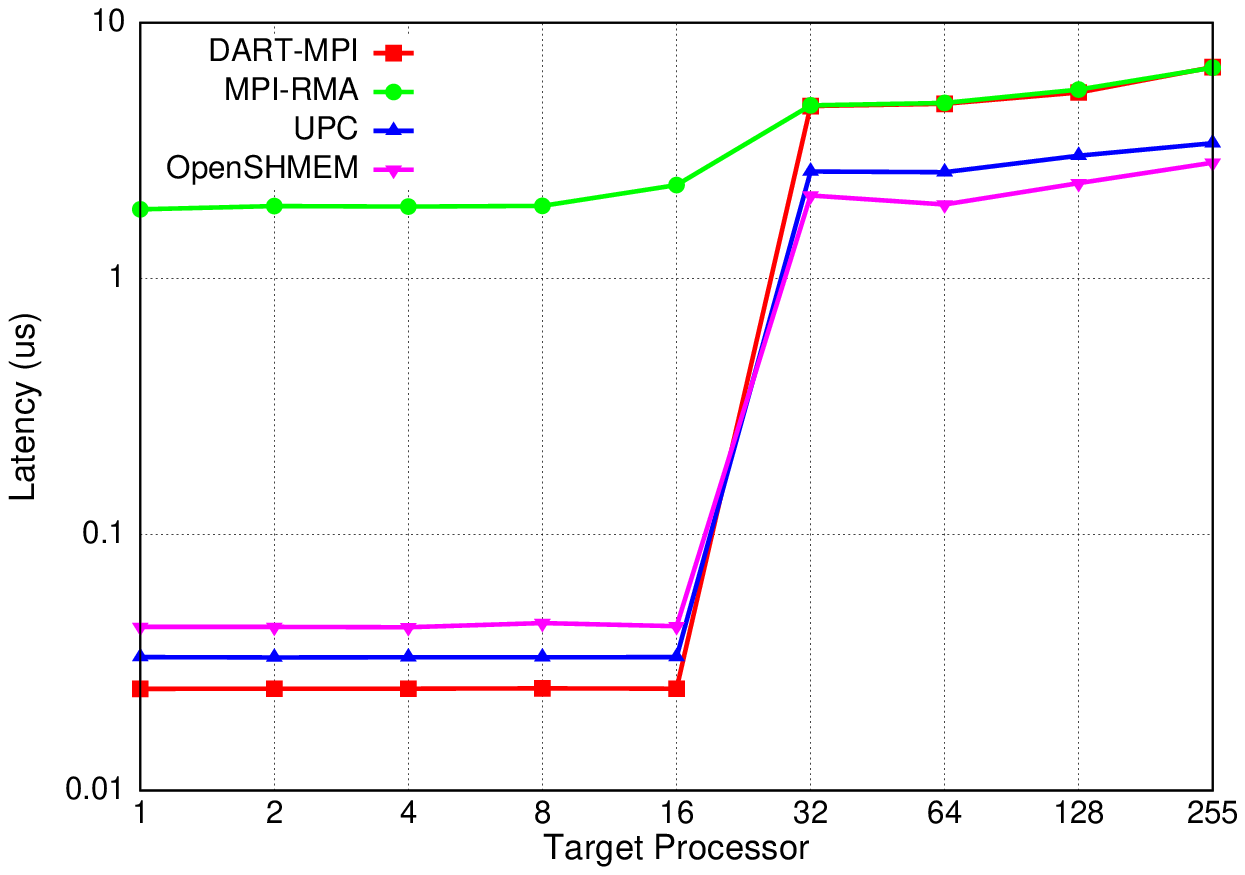}}
\subfloat[Blocking put (1 Mb)]{\includegraphics[width=0.48\textwidth,height=0.21\textheight]{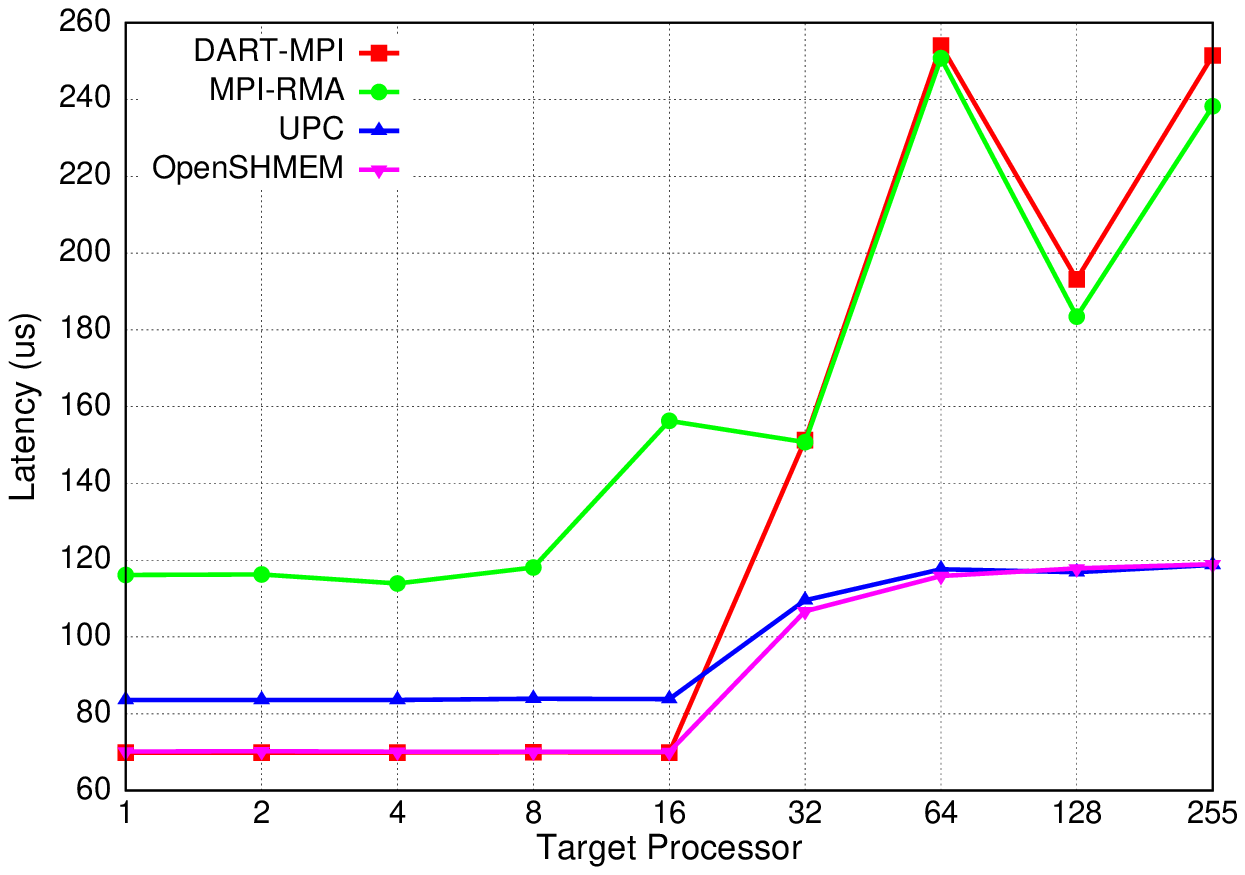}}
\end{center} 
\caption{Blocking put latency as a function of logically increasing distance between two involved ranks/units on 256 PEs}
\label{scalability_put}
\end{figure}
\begin{figure}[tbp]
\begin{center}
\subfloat[Blocking get (8 bytes)]{\includegraphics[width=0.48\textwidth,height=0.21\textheight]{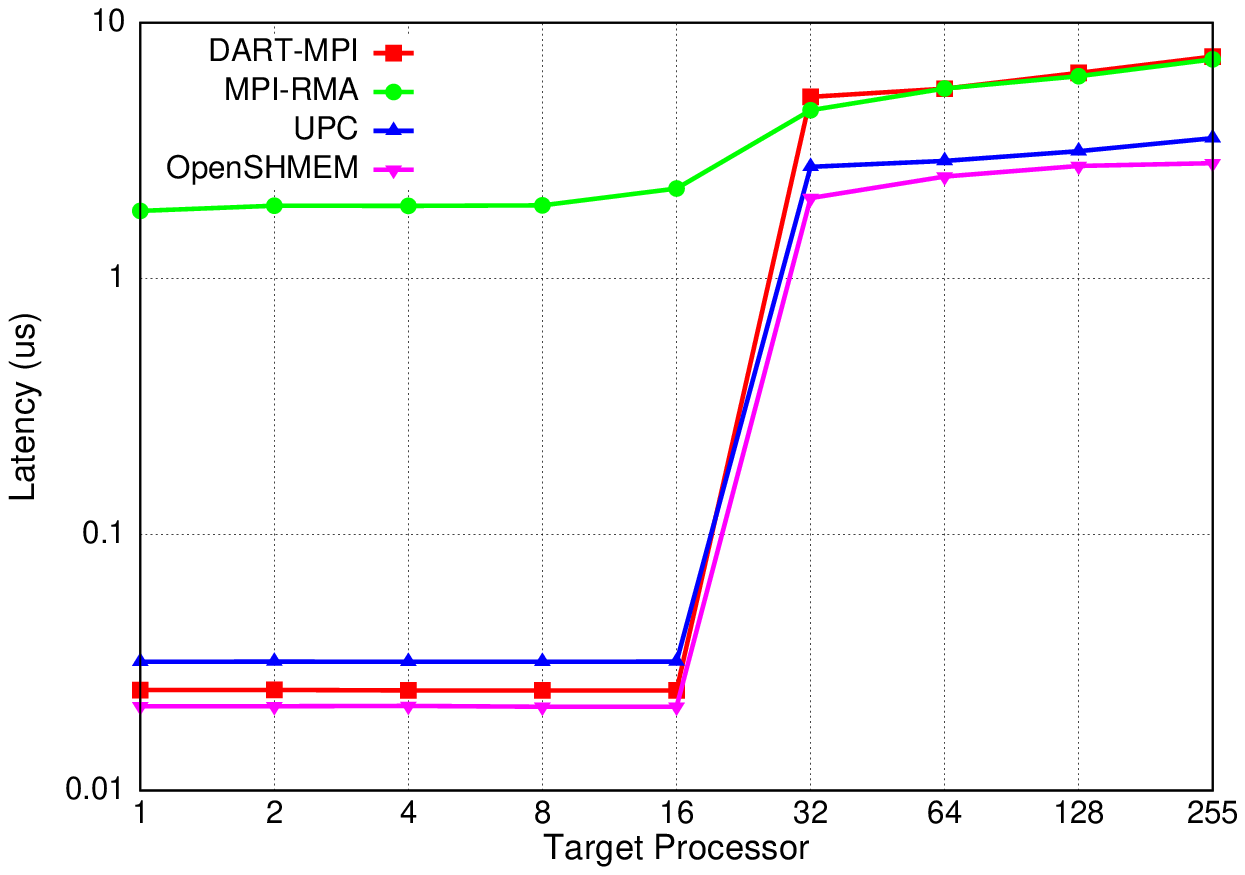}}
\subfloat[Blocking get (1 Mb)]{\includegraphics[width=0.48\textwidth,height=0.21\textheight]{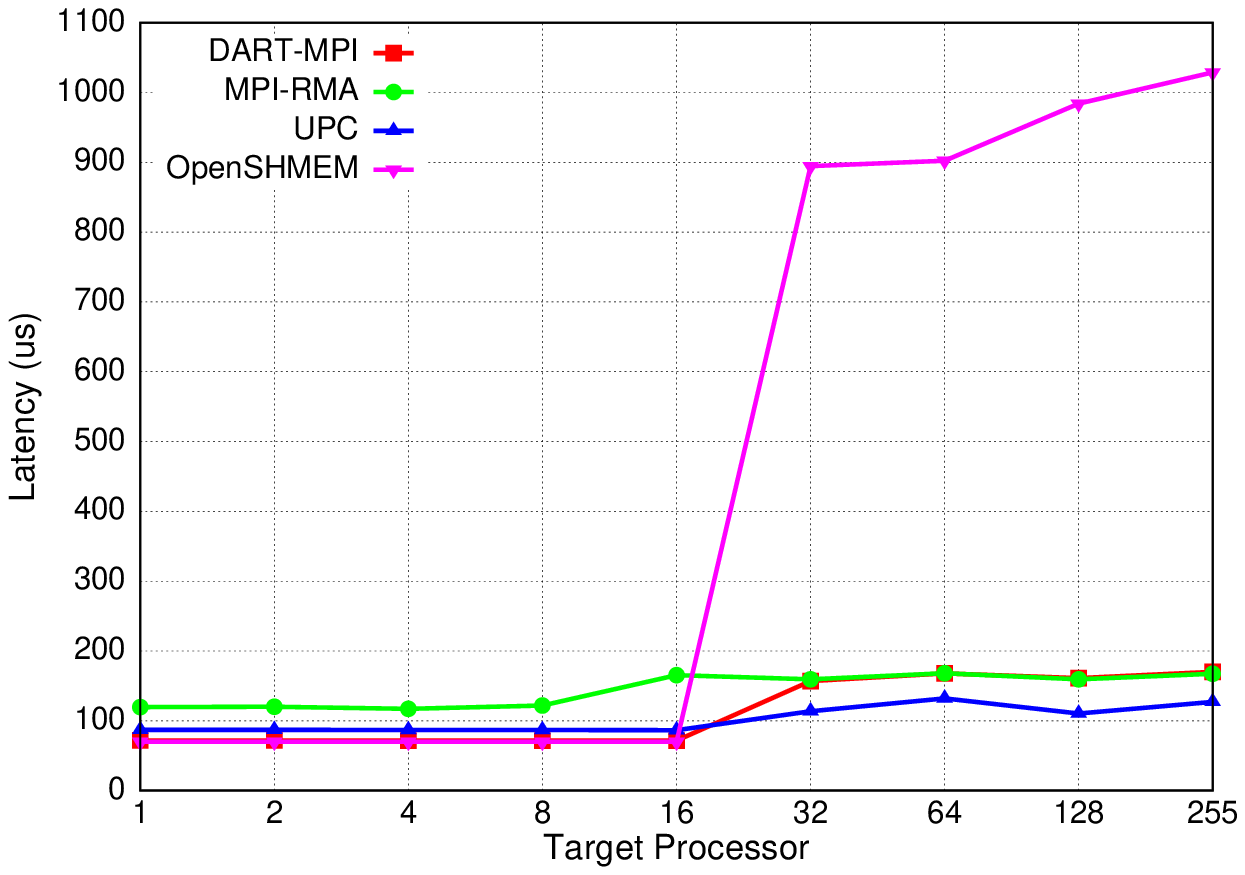} \label{shmemget}}
\end{center} 
\caption{Blocking get latency as a function of logically increasing distance between two involved ranks/units on 256 PEs}
\label{scalability_get}
\end{figure}

In the following, we evaluate the performance of DART-MPI using 
a set of benchmarks which includes low-level communication and application benchmarks.
All the benchmarks are carried out on a Cray XC40 system named Hornet. 
Each compute node features two Intel Haswell E5-2680v3 2.5GHZ processors and consists of
24 cores.
The different compute nodes are interconnected through a Cray Aries
network using Dragonfly topology. They use the Cray-MPI implementation
of MPI-3.

Foremost, we are interested in the evaluation of the performance advantage of
our DART-MPI, using \MPI3 shared-memory and RMA, over native \MPI3
RMA. As shown in a previous paper~\cite{dart-mpi}, the
difference in performance of DART-MPI and MPI-3 RMA operations for non-local
transfers is negligible. In that sense, MPI-3 can be seen as a proxy
for the old DART-MPI. We will thus not show the latter explicitly in this paper.
In addition, we compare \mbox{DART-MPI} with two important PGAS implementations: UPC and OpenSHMEM, 
which are both fully implemented and tuned on the Cray XC40 system. 
In all cases we use the Cray compiler, which also supports UPC
(through the compiler flag \textit{-h upc} ) and OpenSHMEM (as a
library). 
All low-level communication benchmarks are averaged over $10000$ executions.
We do not show the error bars in the following figures, as these
are always small and would only confuse the plots.

\subsection{Low-level Communication Benchmarks}
In this section, we assess the raw communication performance based on the OSU Micro Benchmark\cite{osu_benchmark}.
Firstly, we test the average latencies of the blocking operations of \mbox{DART-MPI} 
as well as the counterparts of MPI (with passive target communication calls), UPC and OpenSHMEM\cite{openshmem} only in the case of intra-node (communication within one node).
Secondly, we evaluate how the blocking put and get operations perform when increasing
logical distance between two involved processes for DART-MPI, MPI, UPC and OpenSHMEM.

Figure \ref{latency} shows the average latency of intra-node blocking
put and get operations for message size ranging from $2^0$ to
$2^{21}$. In all cases the latency roughly keeps constant for small
messages (here $< 1024$ byte). Beyond that the completion time is
dominated by the actual message transfer time and basically grows
linearly with the message size as expected. Noticeably, the curves for
UPC, OpenSHMEM and DART-MPI are very close to each other. For small
messages, native MPI performs more than 10 times slower than the other three
models. This fully illustrates that the overhead of MPI one-sided
operations is relatively high compared to that of direct load/store operations
when data movements happen within one node.

A careful comparison shows, that DART-MPI always performs better for
blocking put operations than UPC (by about 20\%) and OpenSHMEM (by
about 40\%), although such advantage becomes negligible as the message
size increases. For blocking get operations, the variance between
them is much lower in absolute terms, but the trend of curves seems to
suggest that DART-MPI (and to a lesser extend, also UPC) performs slightly
slower than OpenSHMEM.

Next, we evaluate the performance of the blocking RMA operations as a
function of logical distance between source and target. We send
messages of fixed size from process 0 to target processes varying from
1 to 255. Note that the job consists of 256 ranks/units in
total, which corresponds to 11 nodes on Hornet. Figures
\ref{scalability_put} and \ref{scalability_get} show the performance
of blocking put and get operations, respectively, for the short
message size of 8 bytes and the long message size of 1Mb as a function
of logically increasing distance between the origin and target.

As expected the latency remains constant for message transfers within
one node. However, at a logical distance between 16 and 32, i.e., when leaving
one node and targeting the second one, the latency goes up significantly in
all cases except for native MPI, as the overhead of native MPI intra-node data transfers
is relatively high to begin with (as reported above). The
curves for DART-MPI nearly sit above those for native MPI for inter-node data transfers.
This is expected since DART-MPI falls back on
MPI when communicating across nodes. Both, however, perform in general slightly worse than
UPC and OpenSHMEM in latency for inter-node communications. An exception occurs when executing the
OpenSHMEM blocking get operation when transferring large messages; it
performs 5 to 10 times worse than the other three models in latency. We do not have
an explanation for such behavior, but the full data set we have seems
to suggest that OpenSHMEM blocking get operations show relatively poor 
performance for large messages.

\subsection{Application Benchmarks}
In this section we present the results of two application benchmarks,
namely Random Access and a stencil code kernel. All benchmarks were run on up to
1024 cores, i.e., 45 nodes on Hornet.  

\begin{figure}[tbp]
\begin{center}
\includegraphics[scale=0.6]{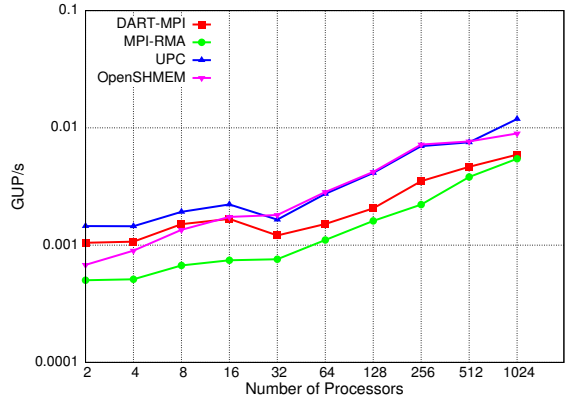}
\end{center} 
\caption{Random Access performance comparison}
\label{ra}
\end{figure}
\begin{figure}[tbp]
\begin{center}
\subfloat[$64 \times 64$]{\includegraphics[width=0.48\textwidth,height=0.21\textheight]{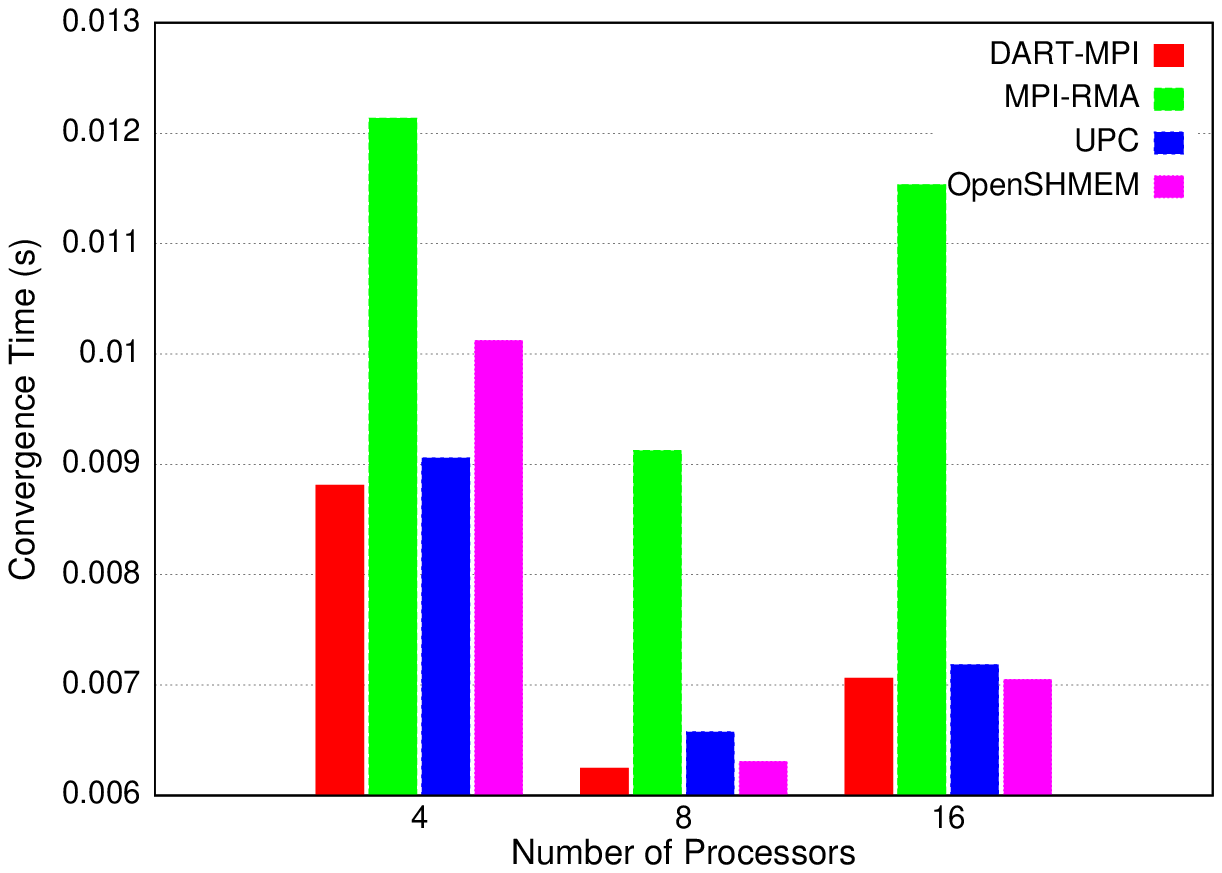} \label{smsg}}
\subfloat[$1024 \times 1024$]{\includegraphics[width=0.48\textwidth,height=0.21\textheight]{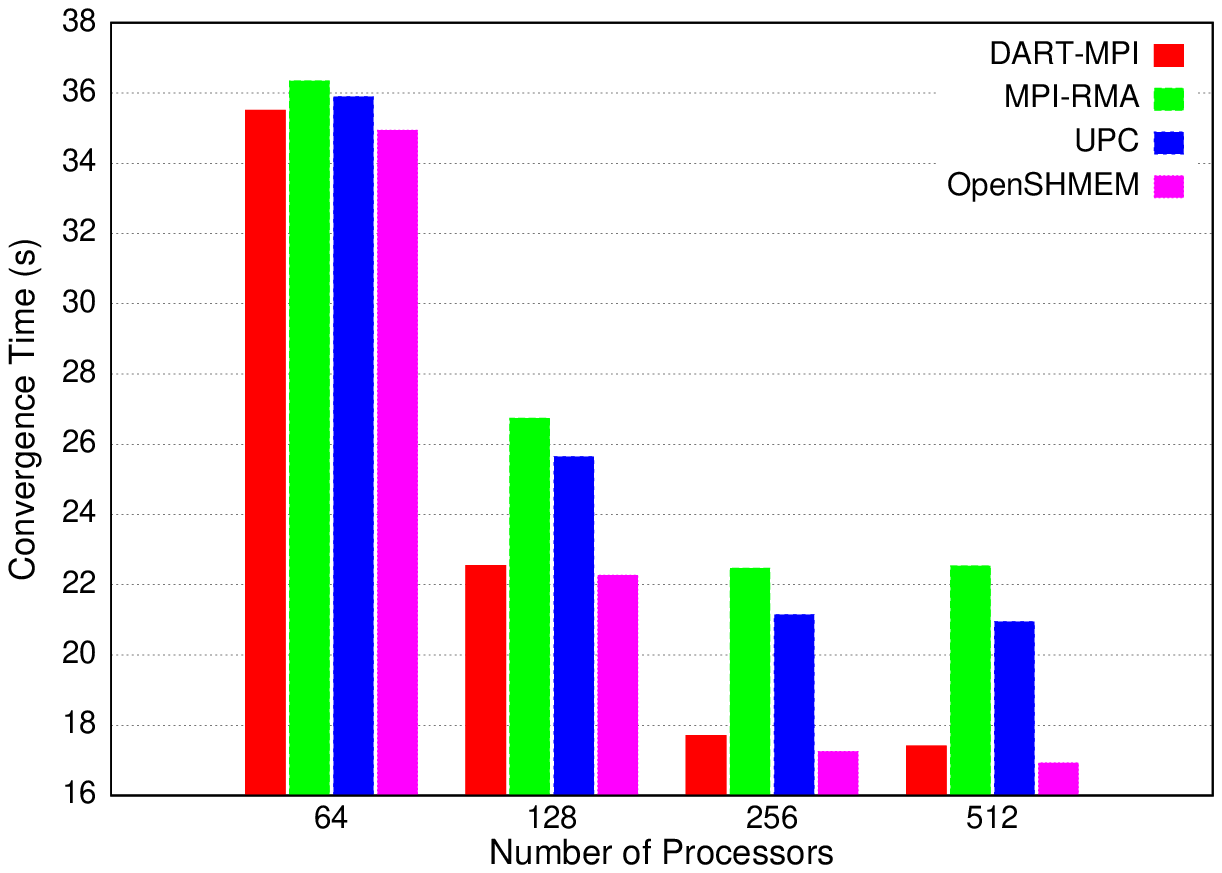} \label{mmsg}} 
\end{center} 
\caption{Five-point stencil performance comparison}
\label{2DPoisson}
\end{figure}

\noindent \textbf{Random Access:} The Random Access (RA) benchmark\cite{ra_rule} is one of the HPC
Challenge benchmarks developed for the HPCS program.  
It consists of concurrent, atomic updates of random elements of a
distributed array by all ranks~\cite{shmem_ra}. 
The general performance metric is
giga-updates per second (GUPs). The messages involved are very
small, i.e., 8 bytes.  Figure \ref{ra} shows the performance in terms
of GUPs
for the DART-MPI, native MPI, UPC and OpenSHMEM versions of 
the RA benchmark for
the number of processes varying from 2 to 1024. 
Interestingly, DART-MPI, UPC and OpenSHMEM achieve similar performance.
The performance of the native MPI version is relatively poor in all cases, and the
performance of the DART-MPI version suffers at large number of ranks due to the
underlying MPI.

The relative performance evaluation of the DART-MPI, UPC and OpenSHMEM
versions is complex stemming from the fact, that the blocking get, put
and atomic operations are involved.
\mbox{DART-MPI} performs slightly better than OpenSHMEM when RA runs
on a single node.  However, we can see there is a clear gap
between the performance of \mbox{DART-MPI} and that of UPC and OpenSHMEM when the
application is carried out across nodes. This is due to the fact that
the amount of the \mbox{inter-node} remote accesses increases as
the growing of the running nodes. The \mbox{inter-node}
communication time performance of \mbox{DART-MPI} is poor relative to
that of UPC and OpenSHMEM for smaller messages (e.g., 8 bytes), as
obvious from Fig. \ref{scalability_put} and
\ref{scalability_get}. 
In addition, the atomic operation contributes
partly to such performance gap between the \mbox{DART-MPI}, UPC and
OpenSHMEM versions, respectively. Noticeably, 
although the increase in the number of inter-node remote accesses
exacerbates the performance of DART-MPI, DART-MPI can
still perform better than native MPI, which has to do with the fraction of memory sharing 
for the intra-node data movements.

\noindent \textbf{Five-Point 2D Stencil Computation:} 
This kernel computes the 2D Poisson equation by applying a \mbox{five-point} stencil on a square grid,
and solving in an iterative way with the \mbox{Gauss-Seidel} method. 
The grid of $N \times N$ elements is decomposed evenly by rows among \textit{numprocs} distributed
processes. Each element holds a \mbox{4-byte} floating point number.
The kernel uses extra halo zones to exchange boundary elements between neighbors,
A total of $4 \times N \times (2 \times \textit{numprocs} - 2)$  bytes
of data per iteration is transmitted using blocking put operations. 
With those halo data, all the inner grid cells can get updated
successfully. We run the stencil kernel until convergence of
solution. The time recorded in the benchmark includes the execution
time of the  Gauss-Seidel solver (local computation part) and
communication time for halo exchange.

We run the five-point stencil benchmark for \mbox{DART-MPI}, MPI, UPC and OpenSHMEM versions 
on the grids of $64 \times 64$ elements and $1024 \times 1024$
elements respectively. Figure \ref{2DPoisson}\protect\subref{smsg} shows comparison results of a $64 \times 64$
grid distributed across 4, 8 and 16 processes on a single node. We can
see that the \mbox{DART-MPI}
version always performs slightly better than the UPC version,
when all the data movement happen within a single node. In addition,
\mbox{DART-MPI}, UPC and OpenSHMEM outperform native MPI by 
$\sim 35\%$ for 16 processes.

The performance of the \mbox{DART-MPI} version degrades when there are data movements across nodes. 
Figure \ref{2DPoisson}\protect\subref{mmsg} shows benchmark results
of a $1024 \times 1024$ grid for 64, 128, 256 and 512 processes. The convergence time
of the DART-MPI and OpenSHMEM versions decreases as the 
number of processes involved is increased, which suggests that DART-MPI and OpenSHMEM 
are more scalable than native MPI and UPC from the perspective of this 
benchmark.


\section {Conclusions}
\label {conclusion}
DART-MPI is the runtime system for the PGAS-like C++ template library
DASH and built on top of \MPI3
one-sided communication primitives. In this paper we present an
optimized design of DART-MPI which uses the new \MPI3 shared-memory
extension for intra-node communications. In essence, we nest \MPI3
shared-memory windows inside RMA windows to do direct load/store
operations for intra-node transfers, and
\MPI3 one-sided communication operations on the RMA windows for
inter-node transfers. 

We expect that this optimization will improve the performance of
DART-MPI for intra-node communication. To verify this claim, we run
three classes of benchmarks, namely low-level put/get benchmarks, a
Random Access benchmark and a stencil application kernel
on the Cray XC40 system. We evaluate 
the performance of DART-MPI against that of native MPI. In addition, we compare DART-MPI to
OpenSHMEM and UPC as two other PGAS-like programming
models. The results of the
evaluation demonstrate, 
first, that DART-MPI performs significantly
faster than MPI RMA when messages are transmitted within a single
node, i.e., that our optimization of DART-MPI leads to a better intra-node 
communication performance, second, that the comparison to OpenSHMEM and UPC show that
the performance improvement that is brought by our optimization makes DART-MPI
comparable with UPC and OpenSHMEM. Additionally, our performance evaluation also shows, that for some
relevant operations -- especially the inter-node RMA operations --
DART-MPI still performs slower than the alternative PGAS approaches.

In this paper, we have only considered blocking RMA put and get
operations. The current design of DART does not include an asynchronous
progress engine, and therefore relies on other parts of the software
stack to do progress as necessary for non-blocking operations. In
particular, we rely on  MPI for non-blocking RMA operations and thus
see no benefit for non-blocking DART operations. An asynchronous
progress engine which allows optimization of non-blocking intra-node
transfers is a subject of further research.

\section*{Acknowledgments}
The authors would like to thank Karl F\"urlinger for fruitful discussion on the DASH runtime
design. 
We gratefully acknowledge funding by the German Research
Foundation (DFG) through the German Priority Programme 1648 Software
for Exascale Computing (SPPEXA).
\bibliographystyle{splncs03}
\bibliography{paper}

\begin{thebibliography}{10}
\providecommand{\url}[1]{\texttt{#1}}
\providecommand{\urlprefix}{URL }

\bibitem{techgasnet}
Bonachea, D., Jeong, J.: {GASNet: A Portable High-Performance Communication
  Layer for Global Address-Space Languages}. Tech. rep., CS258 Parallel
  Computer Architecture Project (2002)

\bibitem{upc}
Carlson, W., Draper, J., Culler, D., Yelick, K., Brooks, E., Warren., K.:
  {Introduction to UPC and Language Specification}. Tech. Rep. CCS-TR-99-157,
  IDA Center for Computing Sciences (1999)

\bibitem{Dinan}
Dinan, J., Balaji, P., Buntinas, D., Goodell, D., Gropp, W., Thakur, R.: {An
  implementation and evaluation of the MPI 3.0 one-sided communication
  interface.} In: {Preprint ANL/MCS-P4014-0113 }. IEEE Computer Society (2013)

\bibitem{dash}
F{\"u}rlinger, K., Glass, C.W., Gracia, J., Kn{\"u}pfer, A., Tao, J.,
  H{\"u}nich, D., Idrees, K., Maiterth, M., Mhedheb, Y., Zhou, H.: {{DASH:}
  Data Structures and Algorithms with Support for Hierarchical Locality}. In:
  {Euro-Par 2014: Parallel Processing Workshops - Euro-Par 2014 International
  Workshops, Porto, Portugal, August 25-26, 2014, Revised Selected Papers, Part
  {II}}. pp. 542--552 (2014),
  \url{http://dx.doi.org/10.1007/978-3-319-14313-2_46;
  http://dblp.uni-trier.de/rec/bib/conf/europar/FurlingerGGKTHIMMZ14}

\bibitem{Hoefler}
Hoefler, T., Dinan, J., Thakur, R., Barrett, B., Balaji, P., Gropp, W.,
  Underwood, K.: {Remote Memory Access Programming in MPI-3}. Tech. rep.,
  Argonne National Laboratory (2013)

\bibitem{mpi22}
{MPI Forum}: {\textsf{MPI}: A Message-Passing Interface Standard. Version 2.2}
  (September 4th 2009), available at: \url{http://www.mpi-forum.org} (Dec.
  2009)

\bibitem{mpi3}
{MPI Forum}: {\textsf{MPI}: A Message-Passing Interface Standard. Version 3.0}
  (September 21st 2012), available at: \url{http://www.mpi-forum.org} (Sept.
  2012)

\bibitem{armci}
Nieplocha, J., Carpenter, B.: {ARMCI: A portable remote memory copy library for
  distributed array libraries and compiler run-time systems.} Tech. rep. (1999)

\bibitem{ga}
Nieplocha, J., Harrison, R.J., Littleeld, R.J.: {Global arrays: A nonuniform
  memory access programming model for high-performance computers}. Journal of
  Supercomputing  10,  169--189 (1996)

\bibitem{osu_benchmark}
{OSU Micro-Benchmarks}. [online] (2014),
  http://mvapich.cse.ohio-state.edu/benchmarks/

\bibitem{pgas}
{Partitioned Global Address Space}. [Online] (2014), http://www.pgas.org/

\bibitem{openshmem}
Poole, S., Hernandez, O., Kuehn, J., Shipman, G., Curtis, A., Feind, K.:
  {OpenSHMEM - Toward a Unified RMA Model}. In: Padua, D. (ed.) {Encyclopedia
  of Parallel Computing}, pp. 1379--1391. Springer US (2011)

\bibitem{dynamic}
Potluri, S., Sur, S., Bureddy, D., Panda, D.K.: {Design and Implementation of
  Key Proposed MPI-3 One-Sided Communication Semantics on InfiniBand.} In:
  Cotronis, Y., Danalis, A., Nikolopoulos, D.S., Dongarra, J. (eds.) {EuroMPI}.
  {Lecture Notes in Computer Science}, vol. 6960, pp. 321--324. Springer
  (2011),
  \url{http://dblp.uni-trier.de/db/conf/pvm/eurompi2011.html#PotluriSBP11;
  http://dx.doi.org/10.1007/978-3-642-24449-0_38;
  http://www.bibsonomy.org/bibtex/2b8d79bab12610e243ae48eceb2905b91/dblp}

\bibitem{ra_rule}
{RandomAccess GUPS (Giga Updates Per Second)}. [online] (2013),
  http://icl.cs.utk.edu/projectsfiles/hpcc/RandomAccess/

\bibitem{shmem_ra}
Shamis, P., Venkata, M.G., Poole, S.W., Welch, A., Curtis., T.: {Designing a
  High Performance OpenSHMEM Implementation Using Universal Common
  Communication Substrate as a Communication Middleware}. In: {OpenSHMEM and
  Related Technologies. Experiences, Implementations, and Tools}. pp. 1--13
  (March 4-6 2014)

\bibitem{async}
Vaidyanathan, K., Chai, L., Huang, W., Panda, D.K.: {Efficient Asynchronous
  Memory Copy Operations on Multi-Core Systems and I/OAT}. In: {IEEE
  International Conference on Cluster Computing} (2007)

\bibitem{dart-mpi}
Zhou, H., Mhedheb, Y., Idrees, K., Glass, C.W., Gracia, J., F{\"u}rlinger, K.,
  Tao, J.: {DART-MPI: An MPI-based Implementation of a PGAS Runtime System.}
  In: {PGAS'14} (Oct 06-10 2014),
  \url{http://dx.doi.org/10.1145/2676870.2676875},
  http://dx.doi.org/10.1145/2676870.2676875

\end{thebibliography}
\end{document}